# PypeIt: The Python Spectroscopic Data Reduction Pipeline

**J. Xavier Prochaska**[1, 2], **Joseph F. Hennawi**[3], **Kyle B. Westfall**[4], **Ryan J. Cooke**[5], **Feige Wang**[2, 6], **Tiffany Hsyu**[1], **Frederick B. Davies**[2, 7], and **Emanuele Paolo Farina**[2, 8]

**1** University of California, Santa Cruz **2** Kavli Institute for the Physics and Mathematics of the Universe **3** University of California, Santa Barbara **4** University of California Observatories **5** Durham University, UK **6** Steward Observatory, University of Arizona **7** Lawrence Berkeley National Laboratory **8** Max Planck Institut f"ur Astrophysik

## Summary

PypeIt is a Python package for semi-automated reduction of astronomical, spectroscopic data. Its algorithms build on decades-long development of previous data reduction pipelines by the developers (Bernstein, Burles, & Prochaska, 2015; Bochanski et al., 2009). The reduction procedure - including a complete list of the input parameters and available functionality - is provided as online documentation hosted by Read the Docs, which is regularly updated. In what follows, we provide a brief description of the algorithms, but refer the interested reader to the online documentation for the most up-to-date information.

Release v1.0.3 serves the following spectrographs: Gemini/GNIRS, Gemini/GMOS, Gemini/FLAMINGOS 2, Lick/Kast, Magellan/MagE, Magellan/Fire, MDM/OSMOS, Keck/DEIMOS (600ZD, 830G, 1200G), Keck/LRIS, Keck/MOSFIRE (Y, J, K gratings tested), Keck/NIRES, Keck/NIRSPEC (low-dispersion), LBT/Luci-I, Luci-II, LBT/MODS (beta), NOT/ALFOSC (grism4), VLT/X-Shooter (VIS, NIR), VLT/FORS2 (300I, 300V).

This v1.0 release of `PypeIt` is designed to be used by both advanced spectroscopists with prior data reduction expertise and astronomers with no prior experience of data reduction. It is highly configurable and designed to be applied to any standard slit-imaging spectrograph, and can accomodate longslit, multislit, as well as cross-dispersed echelle spectra. It has already enabled several scientific publications (Eilers, Hennawi, & Davies, 2018; Eilers et al., 2020; Hsyu, Cooke, Prochaska, & Bolte, 2018; Wang et al., 2020; Yang, Wang, Fan, Hennawi, Davies, Yue, Banados, et al., 2020; Yang, Wang, Fan, Hennawi, Davies, Yue, Eilers, et al., 2020).

In order to successfully reduce your data with `PypeIt`, we recommend that you obtain the following calibration frames as a minimum, using these guidelines:

(i) Flat frames (at least 1 frame, and ideally more than 5) to be used for slit/order edge tracing and relative pixel efficiency correction. These frames should be acquired with the same slit width and setup as your science frames.

(ii) Arc frames (at least 1 frame, and ideally ~3 to improve S/N of weak lines) to be used for wavelength calibration. Please see the online `PypeIt` documentation for suggestions about the choice of lamps that you should use for your instrument/setup. These frames should be acquired with the same slit width and setup as your science frames. For the near-IR PypeIt uses sky lines for wavelength calibration so arcs are not required.

Depending on your science goal, and the instrument being used, you may also require the following optional frames:

(iii) Bias or Dark frames (ideally 10 frames) should be acquired with a 0 second exposure with the shutter closed, while dark frames (ideally 3 frames) should be acquired with the shutter closed, ideally with an exposure time of equal duration to your science frames.

(iv) Twilight sky frames (ideally 3 frames) using a slit width of equal size to your science frames. These frames are used to construct slit illumination functions for situations where the internal or dome flats: either produce a different illumination function than the sky (i.e. often for internals) or suffer from insufficient counts (often an issue in the blue).

(v) Pixel flat frames (ideally more than 3) which are taken with the detectors uniformly illuminated (or, with a slit width larger than that taken with the science frames).

(vi) Standard star frames (at least 1 frame, ideally 3) to calibrate your spectrum (depending on your science goal, this may be a flux, telluric, color, or velocity standard).

After the creation of a custom input/configuration file, the pipeline runs end-to-end to convert raw spectroscopic images into calibrated, science-ready spectra. In what follows, we describe several key steps of the data reduction procedure:

(1) The pipeline automatically characterises the raw input frames based on header information. We have also developed a SPectral Image Typing (SPIT) tool (Jankov & Prochaska, 2018) to classify images based on the pixel data. The output of this classification procedure, is an input `PypeIt` file that allows the user to specify the parameters of their reduction, and manually update the classification of the input raw images.

(2) The reduction procedure consists of a script that automatically applies a series of algorithms to the raw data frames. All raw images are first overscan subtracted, and optionally corrected for the bias and dark current. A bad pixel mask is generated internally, or can be constructed using the bias or dark frames, if available. All frames of the same type are robustly combined to construct master calibration frames.

(3) The edges of the slit are typically traced using frames where the slit is uniformly illuminated by either a halogen lamp or a spectrum of the twilight sky (especially for blue setups).

(4) A master arc frame (or the science data in the near-IR) is used for wavelength calibration and to generate a map of the wavelength solution across the entire detector. This accounts for the spectral tilt across the slit. `PypeIt` contains an archive of wavelength solutions that are used to determine the wavelength solution of your data. If the automated wavelength calibration technique does not succeed, `PypeIt` includes a script with a graphical user interface that allows the user to calibrate their spectra. If you perform a wavelength calibration, we kindly request that you share your solution, and we will make this available to the community.

(5) `PypeIt` generates a 2D model of the flat frame, which is used to construct a pixelflat, and to determine the spatial slit profile. Throughout the reduction procedure, the slit trace image is also used to calculate the spatial flexure of each frame relative to the master flat frame.

(6) The above calibrations are applied to every science and standard star frame. `PypeIt` jointly performs the object extraction and b-spline @kelson2003 sky subtraction. A two-dimensional model of the sky is first constructed using the spectral tilt map, including a robust fit to separate the signal of the science target from the sky background emission.

This sky model is locally refined around the science target during spectrum extraction. The algorithm we have developed for `PypeIt` achieves Poisson limited sky-subtraction (see Figure Figure 1). `PypeIt` then performs a boxcar and an optimal extraction to generate 1D science spectra. The final output of this procedure is a series of fully reduced one- and two-dimensional spectra.

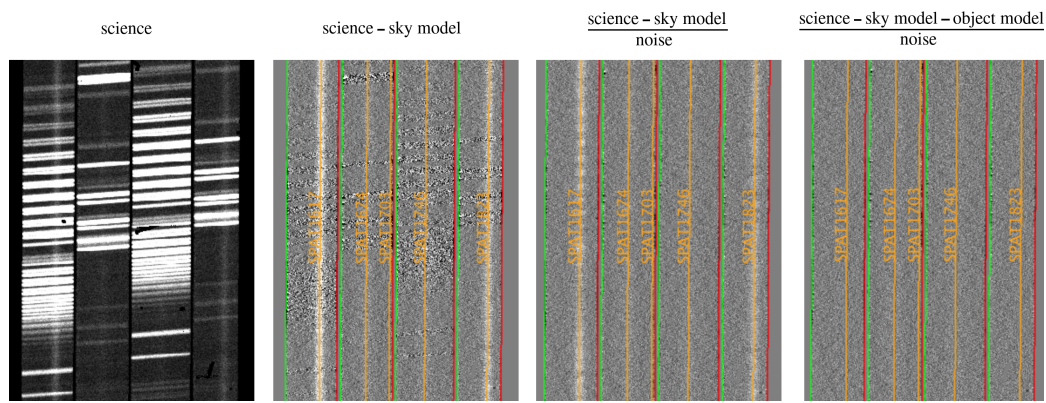

**Figure 1:** PypeIt sky subtraction reaches the Poisson limit. This image shows several two-dimensional Keck/DEIMOS spectra (1200G grating). From left to right: (1) Raw science frame, (2) Processed science frame after a model of the sky emission has been removed, (3) Same as previous, but relative to the noise to highlight object spectrum and sky residuals (4) Same as previous, but a model of the object is also removed. This produces a final processed 2D frame at the Poisson limit.

(7) As a final step, the wavelength solution of the one-dimensional extracted spectra are corrected for spectral flexure with reference to the sky emission lines. The wavelength solution can then be transformed to the user-spectified frame of reference (usually heliocentric or barycentric).

Finally, `PypeIt` also includes scripts to flux calibrate and combine multiple one- and two-dimensional exposures, as well as software for performing telluric corrections. `PypeIt` produces a series of calibration-related outputs and includes scripts and automatically generated plots for quality assurance inspection. The final outputs are FITS files with rigid well-documented data models that hold the two-dimensional (includes spatial information) and one-dimensional spectral extractions.

It is our plan to expand `PypeIt` to include the majority of spectrographs on the largest ground-based optical and near-infrared telescopes, ideally with help from the broader community. We are currently working towards implementing the following additional spectrographs: Keck/DEIMOS (all gratings) Keck/KCWI, Keck/MOSFIRE (all setups), Keck/NIRSPEC (new detector + high resolution), Keck/ESI, Keck/HIRES, Magellan/IMACS, MMT/BinoSpec, VLT/UVES. We are also open to receiving requests to support additional spectroscopic instrumentation. Join us on GitHub. We ask those interested in developing and enhancing PypeIt to agree to our code of conduct.

## Acknowledgements

We acknowledge intellectual contributions from Scott Burles, Rob Simcoe, and David Schlegel.

`PypeIt` has been financially supported by the University of California Observatories. J. F. H. also acknowledges support from the University of California, Santa Barbara. During work on `PypeIt`, R. J. C. was supported by a Royal Society University Research Fellowship, and acknowledges support from STFC (ST/P000541/1, ST/T000244/1).